\documentclass{aa}
\usepackage{psfig}
\begin{document}

   \thesaurus{06     % A&A Section 6: Form. struct. and evolut. of stars
              (02.04.1;  % Dense matter,
               08.05.3;  % Stars: evolution,
               08.09.3;  % Stars: interiors,    
               08.14.1;  % Stars: neutron,      
               08.18.1)} % Stars: rotation.     
   \title{Deconfinement transition in rotating compact stars}

%   \subtitle{}

   \author{E.~Chubarian \inst{1}
        \and H. Grigorian \inst{1,2}
        \and G.~Poghosyan \inst{1}
        \and D. Blaschke \inst{2}
          }

   \offprints{D. Blaschke}

   \institute{ Department of Physics, Yerevan State University, Alex
        Manoogian Str. 1, 375025 Yerevan, Armenia
        \and Fachbereich Physik, Universit\"at Rostock,
        Universit\"atsplatz 1, D--18051 Rostock, Germany\\
         email: blaschke@darss.mpg.uni-rostock.de 
%       \and Kernfysisch Versneller Instituut, NL-7947 AA Groningen, 
%       The Netherlands
%             \thanks{The university of heaven temporarily does not
%                     accept e-mails}
             }

   \date{Received 1 February 2000/ Accepted 15 February 2000}

   \maketitle

   \begin{abstract}
Using the formalism of general relativity for axially symmetric gravitational 
fields and their sources - rotating compact stars -
a perturbation theory with respect to angular velocity is developed and  
physical quantities such as mass, shape, momentum of inertia and total energy 
of the star are defined. 
The change of the internal structure 
of the star due to rotation has been investigated 
and the different contributions to the 
moment of inertia have been evaluated separately.
Numerical solutions have been performed using a two-flavor model equation of 
state describing the deconfinement phase transition as constrained by the 
conservation of total baryon number and electric charge.
During the spin down evolution of the rotating neutron star, below critical 
values of angular velocity a quark matter core can appear which might be 
detected as a characteristic signal in the pulsar timing. 
Within the spin-down scenario due to magnetic dipole radiation it is shown that
the deviation of the braking index from $n=3$ could signal not only the 
occurrence but also the size of a quark core in the pulsar.
A new scenario is proposed where, due to mass accretion onto the rapidly 
rotating compact star, a spin-down to spin-up transition might signal a 
deconfinement transition in its interior. 

   \end{abstract}

%
%________________________________________________________________

\section{Introduction}
In many recent astrophysical applications of the theory of dense matter it is 
necessary to investigate the properties of rapidly rotating compact objects 
within general relativity theory. 
The reason for this development is the hope that changes in the internal 
structure of the dense matter, e.g. during phase transitions, could have 
observable consequences for the dynamics of the rotational behavior of these 
objects. Particular examples are the observations of glitches and postglitch 
relaxation in pulsars which are discussed as signals for superfluidity in 
nuclear matter (Pines \& Ali Alpar 1992) and the suggestion that the braking 
index is remarkably enhanced when a quark matter core occurs in the centre of 
a pulsar during its spin-down evolution (Glendenning et al. 1997). 
Further constraints for the nuclear equation of state come from the 
observation of quasi-periodic brightness oscillations (QPO's) in low-mass 
X-ray binaries which entail mass and radius limits for rapidly rotating 
neutron stars, see \cite{rxte,bombaci}.   

The problem of rotation in the general relativity theory was and remains one 
of the central and complicated problems (Glendenning 1997). 
Compared to the modern methods of numerical solutions to this problem 
(Friedman et al. 1986, Salgado et al. 1994) the 
method of perturbation theory (Hartle 1967, Sedrakian \& Chubarian 1968) is a 
physically transparent and systematic approach to the solution of the problem 
of stationary gravitational fields and their sources. 
This approach has been applied successfully in general
relativity as well as in alternative theories of gravitation 
(Grigorian \& Chubarian 1985).

From a practical point of view in the definition of the integral 
characteristics of the astrophysical objects, it is important to analyze the 
asymptotical expansion  of the metric tensor at 
large distances from the stars, to be able to compare the results with 
observational data. 
One can of course introduce the physical parameters of the configuration 
using the symmetry properties of the object and the gravitational field 
by expressing them in terms of conserved quantities. 
In this work we use the definition of the $J_z$ projection of the 
angular momentum of the nonspherical rotating star
(z is the axis of the star's rotation and symmetry of the gravitational field)
as a conserved integral of motion. 
It is a well known integral of the non diagonal element of the energy-momentum
tensor in the frame of spherical coordinates. 

Using the method of perturbation theory, we are going to calculate
the total mass, angular momentum and shape deformation from the 
solution of the gravitational field equations in case of 
hydrodynamical, thermodynamical and chemical equilibrium for a given
total baryon number and angular velocity $\Omega$ of the object. 
According to \cite{chubarian} 
the expansion parameter is of the order of the ratio 
of the rotational and gravitational energy. 
It has been shown 
that, for compact objects in the stationary rotating regime without matter 
flux, the first two terms of the series expansion give a 
sufficiently good approximation (see also Glendenning 1997, Weber 1999).  
 
The evolution of the rotating stars could have different origins and 
scenarii. 
Our aim in this work is to discuss possible signals for a deconfinement 
phase transition during the evolution of a rotating compact object on the basis
of solutions for the $\Omega$ dependence of the moment of inertia. 

\section{Self-consistent set of field equations for stationary rotating stars}
\label{sec:GR}
\subsection{Einstein equations for axial symmetry}

The general form of the metric for an axial symmetric space-time manifold is
\begin{eqnarray}  
\label{metric}
ds^2=e^{\nu} dt^2 - e^{\lambda}dr^2-r^2e^{\mu}[d\theta^2+\sin^2\theta(d\phi+
\omega dt)^2]~~,
\end{eqnarray}
written in a spherical symmetric coordinate system in order to obtain the 
Schwarzschild solution as a limiting case.
This line element is time-translational and axial-rotational invariant; all
metric functions are dependent on the coordinate distance from the
coordinate center $r$ and azimuthal angle $\theta$ between the radius vector
and the axis of symmetry.

Reversal symmetry of the time and polar angle $\phi$ require that all
metric coefficients except $\omega$ must be even functions of the angular
velocity 
\begin{equation}  
\label{omega}
\Omega=\frac{d\phi}{dt}
\end{equation}
of the star, the gravitational field of which is described by the Eq. (\ref
{metric}). The physical characteristics of the rotating object depend on the
centrifugal forces in the local inertial frame of the observer. In general
relativity, due to the Lense-Thirring law, rotational effects are described by 
$\bar \omega$ the difference of the frame dragging frequency $-\omega$ and the
angular velocity $\Omega$ 
\begin{equation}  \label{dromega}
\bar \omega \equiv \Omega+ \omega(r,\theta).
\end{equation}

The energy momentum tensor of stellar matter can be approximated by the
expression of the energy momentum tensor of an ideal fluid 
\begin{equation}  
\label{Energy}
T_{\mu}^{\nu}=(\varepsilon + p)u_{\mu}u^{\nu}-p\delta_{\mu}^{\nu},
\end{equation}
where $u^{\mu}$ is the $4$-velocity of matter, $p$ the pressure and 
$\varepsilon$ the energy density.

We assume that the star due to high viscosity (ignoring the super-fluid
component of the matter) rotates stationarily as a solid body with an
angular velocity $\Omega$ that is independent of the spatial coordinates. 
The time scales for changes in the angular velocity which we will consider in 
our applications are well separated from the relaxation times at which 
hydrodynamical equilibrium is established, so that the assumption of a 
rigid rotator model is justified.

Therefore there are only two non-vanishing components of the velocity 
\begin{eqnarray}  
\label{velocity}
u^{\phi}&=&\Omega~u^t~,  \nonumber \\
u^t&=&1/\sqrt{ e^{\nu}-r^2e^{\mu}\bar \omega^2 sin^2\theta}~.
\end{eqnarray}

The equation of state which we will use for our investigation of the 
deconfinement transition in rotating compact stars will be introduced in 
Sect. \ref{sec:eos}.
Once the energy-momentum tensor (\ref{Energy}) is fixed by the choice of the
equation of state for stellar matter, the unknown metric functions 
$\nu$,$\lambda$, $\mu$, $\bar \omega$ can be determined by the set of 
Einstein field equations for which we use the following four combinations. 

There are three Einstein equations for the determination of the diagonal 
elements of the metric tensor 
\begin{eqnarray}  
\label{einstein}
G_{r}^{r}- G_{t}^{t}&=&8\pi G ( T_{r}^{r}- T_{t}^{t})~,  \nonumber \\
G_{\theta}^{\theta}+ G_{\phi}^{\phi}&=&8\pi G ( T_{\theta}^{\theta}+
T_{\phi}^{\phi})~,  \nonumber \\
G_{\theta}^{r}&=&0~,
\end{eqnarray}
and one for the determination of the non diagonal element 
\begin{equation}  
\label{rote}
G_{\phi}^{t}=8\pi G T_{\phi}^{t}~.
\end{equation}
Here $G$ is the gravitational constant and $G_{\mu}^{\nu}$ the Einstein
tensor.

We use also one equation for the hydrodynamical equilibrium (Euler
equation) 
\begin{equation}  \label{euler}
{H} (r,\theta) \equiv
\int \frac{d p^{\prime}}{p^{\prime}+\varepsilon^{\prime}}
= \frac{1}{2}\ln [u^t(r,\theta)] + {\rm const},
\end{equation}
where the gravitational enthalpy ${H}$ thus introduced is a function
of the energy and/or pressure distribution.

The parameters of the theory are the angular velocity of the rotation 
$\Omega$ and the central energy density $\varepsilon(0)$ of the star 
configuration.

\subsection{Perturbative approach to the solution}
\label{sec:mass} 
The problem of the rotation can be solved iteratively by using a
perturbation expansion of the metric tensor and the physical quantities in a 
Taylor series with respect to the angular velocity. 
As a small parameter for this expansion we use a dimensionless quantity
which is the ratio of the rotational 
energy to the gravitational one for a homogeneous Newtonian star 
$E_{\rm rot}/E_{\rm grav} = (\Omega / \Omega_0)^2$, 
where $ \Omega_0^2 = 4\pi G \rho(0)$  with the mass density 
$\rho(0)$ at the center of the star. 
This expansion gives sufficiently correct solutions already at $O(\Omega^2)$,
since the expansion parameter is limited to values 
$\Omega / {\Omega}_0 \ll 1$ by the condition of 
mechanical stability of the rigid rotation. 
This can easily be seen by considering as an upper limit for attainable 
angular velocities the so called Kepler one $\Omega _K= \sqrt{G M/ R_e^3}$  
with $M$ being the total mass and $R_e$ the equatorial radius.
For homogeneous Newtonian spherical stars 
$\Omega < \Omega _K = {\Omega}_0/\sqrt{3}$ which is fulfilled in particular 
also for the configurations with a deconfinement transition which we are going 
to discuss later in this paper, see Fig. \ref{fig4}.

The expansion of the metric tensor is given by
\begin{equation}
g_{\mu \nu}(r,\theta;\Omega)=\sum_{j=0}^{\infty} 
\left(\frac{\Omega}{\Omega_0}\right)^j g^{(j)}_{\mu \nu}(r,\theta)~.
\end{equation}   
According to the symmetries of the metric coefficients introduced in Eq. 
(\ref{metric}) we have even orders $j=0,2,....$ for the diagonal 
elements\footnote{Notation corresponds to the works of \cite{chubarian}.} 
\begin{eqnarray}
\label{series}
e^{-\lambda (r,\theta;\Omega)} 
&=&e^{-\lambda^{(0)}(r)}[1+(\Omega/{\Omega}_0)^2 f(r,\theta )]+O(\Omega^4)~,
\nonumber\\
e^{\nu (r,\theta;\Omega)} 
&=&e^{\nu^{(0)}(r)}[1+(\Omega/{\Omega}_0)^2 \Phi (r,\theta )]+O(\Omega^4)~, 
\nonumber \\
e^{\mu (r,\theta;\Omega)} 
&=&r^2[1+(\Omega/{\Omega}_0)^2 U(r,\theta )]+O(\Omega^4)~,
\end{eqnarray}
and odd orders only for the frame dragging frequency $\omega $ 
\begin{equation}
\omega(r,\theta;\Omega) 
=\frac{\Omega}{{\Omega}_0}~q(r,\theta)+O(\Omega^3). 
\label{intq}
\end{equation}

The distributions of pressure, energy density and ``enthalpy'' introduced in 
Eq. (\ref{euler}) are also included in the scheme of this perturbation 
expansion
\begin{eqnarray}
\label{euler1}
p(r,\theta;\Omega) 
&=&p^{(0)}(r)+(\Omega/{\Omega}_0)^2 p^{(2)}(r,\theta )+O(\Omega^4)~,  
\nonumber\\
\varepsilon (r,\theta;\Omega) 
&=&\varepsilon^{(0)}(r)+ (\Omega/{\Omega}_0)^2 \varepsilon^{(2)}(r,\theta)
+O(\Omega^4),  \nonumber\\
{H}(r,\theta;\Omega) 
&=&{H}^{(0)}(r)+ (\Omega/{\Omega}_0)^2 H^{(2)}(r,\theta )+O(\Omega^4).
\end{eqnarray}
All functions with superscript ($0$) denote the solution of the static
configuration and therefore they are only functions of the distance from the 
center $r$, the others are the corrections corresponding to the rotation.

This series expansion allows one to transform the Einstein equations into 
a coupled set of equations for the coefficient functions which can be solved 
by recursion.
At zeroth order we recover the nonlinear problem of the static spherically 
symmetric star configuration (Tolman-Oppenheimer-Volkoff equations), 
see the next subsection \ref{ssec:static}.
The first recursion step is to solve Eq. (\ref{rote}) in order to obtain
the dragging frequency in $O(\Omega)$ and to define the moment of inertia for 
the spherically symmetric configuration.
In subsection \ref{ssec:deltaI} we will consider the second order 
contribution in the $\Omega$- expansion (\ref{series}), (\ref{euler1}) 
where the $O(\Omega^2)$ corrections to the moment of inertia can be found.
The next terms in the expansion which are of $O(\Omega^3)$ correspond to 
corrections of the frame dragging frequency and will be neglected since they 
go beyond the approximation scheme adopted in the present paper.

\subsection{Zeroth order: Static spherically symmetric star models}
\label{ssec:static} 

The functions of the spherically symmetric solution in Eqs. (\ref{series}) and 
(\ref{euler1}) can be
found from Eq. (\ref{einstein}) and Eq. (\ref{euler}) in zeroth order of
the $\Omega$- expansion.

They obey the following equation (Tolman-Oppenheimer-Volkoff) 
\begin{equation}  
\label{TOV}
\frac{dp^{(0)}(r)}{dr}=-G [p^{(0)}(r)+\varepsilon^{(0)}(r)] 
\frac{m(r)+4 \pi p^{(0)}(r) r^3}{r[r-2 G m(r)]}~,
\end{equation}
where $m(r)$ is the distribution of accumulated mass 
\begin{equation}  
\label{mass}
m(r)=4\pi \int_0^r \varepsilon^{(0)}(r^{\prime}) r^{\prime 2} dr^{\prime}
\end{equation}
within a sphere of radius $r$. For the gravitational potentials we have 
\begin{eqnarray}  
\label{phs}
\lambda^{(0)}(r)&=&-\ln [1- {2 G m(r)}/{r}] ~,\\
\nu^{(0)}(r)&=&-\lambda^{(0)}(R_0) - 2 G \int^{R_0}_r \frac{m(r^{\prime})+4\pi
p^{(0)}(r^{\prime})r^{\prime 3}}{r^{\prime}[r^{\prime}-2 G m(r^{\prime})]}%
dr^{\prime}~.  
\nonumber \\
\end{eqnarray}
$R_0$ is the spherical radius of the star, which is defined by 
$P^{(0)}(R_0)=0$. 
The set of Eq.(\ref{TOV}) and Eq.(\ref{mass}) fulfils the following
conditions at the center of the configuration: 
$\varepsilon^{(0)}(0)=\varepsilon(0)$ and $m(0)=0$. 
The central energy density $\varepsilon(0)$ is the parameter of the spherical
configuration.
The total mass of the spherically distributed matter in the 
selfconsistent gravitational field is $M_0(\varepsilon^{(0)}(0))=m(R_0)$. 

\subsection{Moment of inertia}
\label{sec:MI}

In the first order of the approximation we are solving 
Eq.(\ref{rote}), where the unknown function  $q(r,\theta )$ defined by
Eq. (\ref{intq})  is independent of the angular velocity. Using the static
solutions  Eqs.(\ref{TOV})-(\ref{phs}), and the
representation of  $q(r,\theta)$ by the series of the Legendre
polynomials,
\begin{equation}
q(r,\theta )=\sum_{m=0}^{\infty}q_m(r)
\frac{dP_{m+1}(\cos \theta)}{d\cos\theta} ~,
\end{equation}
we find the equations for the coefficients $q_m(r)$.
It is proved that  $q(r,\theta )$ is a function of the
distance $r$ only, i.e. that $q_m(r)=0$ for $m>0$ , see 
\cite{hartle,chubarian}.

Let us write down the equations for
$\bar \omega(r)=\Omega(1+q_0(r)/{\Omega}_0)$, 
which is more suitable for the solution of the resulting equation in first 
order 
\begin{equation}
\frac 1{r^4}\frac d{dr}\left[r^4j(r)\frac{d\bar \omega(r) }{dr}\right]
+\frac 4r\frac{dj(r)}{dr}\bar \omega(r)=0~,  
\label{Hartl}
\end{equation}
which corresponds to Ref. \cite{hartle}, where it was obtained using a 
different representation of the metric.
Here, we use the notation $j(r)\equiv e^{-(\nu^{(0)}(r)+\lambda^{(0)}(r))/2}$,
for which outside of configuration $j(r)=1~,~~r>R_0$. 

By definition, the angular momentum of the
star in the case of stationary rotation is a conserved quantity and can be 
expressed in invariant form 
\begin{equation}
J=\int T_\phi ^t\sqrt{-g}dV~,  
\label{moment}
\end{equation}
where $\sqrt{-g}dV$ is the invariant volume and $g=\det||g_{\mu\nu}||$. 
For the case of slow rotation where the shape deformation of the rotating 
star can be neglected and using the definition of the moment of inertia
$I_0(r)=J_0(r)/\Omega$ accumulated in the sphere with radius $r$, we obtain 
from Eq. (\ref{moment})
\begin{eqnarray}  
\label{omoment}
\frac{dI_0(r)}{dr}=\frac{8\pi}{3} r^4[\varepsilon^{(0)}(r)+p^{(0)}(r)]
e^{(-\nu^{(0)}(r)+\lambda^{(0)}(r))/2}\frac{\bar \omega(r)}{\Omega}~.
\nonumber \\
\end{eqnarray}

Using this equation one can reduce the second order differential equation 
(\ref{Hartl}) to the first order one
\begin{equation}
\frac{d\bar \omega(r) }{dr}=\frac{6GJ_0(r)}{r^4j(r)}.  
\label{dmoment}
\end{equation}
and solve (\ref{Hartl}) as a coupled set of first order differential 
equations, one for the moment of inertia (\ref{omoment})  and the
other (\ref{dmoment}) for the frame dragging frequency $\bar \omega(r)$. 

This system of equations is valid inside and outside the matter distribution.
In the center of the configuration $I_0(0)=0$ and 
$\bar \omega(0)=\bar \omega_0$. The finite value $\bar \omega_0$
has to be defined such that the dragging
frequency $\bar \omega (r)$ smoothly joins the outer solution 
\begin{equation}
\bar \omega (r)=\Omega \left(1 -\frac{2G I_0}{r^3}\right).  
\label{dgvac}
\end{equation}
at $r=R_0$,
and approaches $\Omega $ in the limit $r\to\infty$. In the external solution
(\ref{dgvac}) the constant $I_0=I_0(R_0)$ is the total moment of inertia of 
the slowly rotating star and $J_0=I_0 \Omega $ is the corresponding  angular 
momentum. In this order of approximation, $I_0$ is a function of the central
energy density or the total baryon number only. Explicit dependences of the
moment of inertia on the angular velocity occur in the second order of 
approximation.

\subsection{Second order corrections to the moment of inertia}
\label{ssec:deltaI}

Due to the rotation in $\Omega^2 $-approximation the shape of the star is an
ellipsoid, and each of the equal-pressure (isobar) surfaces in the star is
an ellipsoid as well.  
All diagonal elements of the metric and energy-momentum tensors could be
represented as a series expansion in Legendre polynomials 
\begin{equation}
g^{(2)}_{\mu \nu}(r,\theta)=
\sum_{l=0}^{\infty} (g_{\mu \nu})_l(r)P_{l}(\cos \theta)~.
\label{eq:polinom}
\end{equation}
It has been shown that the only non vanishing solutions obeying the 
continuity conditions on the surface are those with $l=0,2$.  

The deformation of the isobaric surfaces due to the rotation can be 
parametrized by the shift 
$R(r,\theta)-r=\Delta (r,\theta )$ 
which describes the deviation from the spheric distribution as a function
of the radius $r$ in the given polar angle $\theta$ and is completely 
determined by 
\begin{equation}
R(r,\theta )=r+\left(\frac{\Omega}{ \Omega_0}\right)^2 
[\Delta_0(r)+\Delta_2(r)P_2(\cos\theta)],  
\label{def}
\end{equation}
since the  expansion coefficients of the deformation $\Delta_l(r)$
can be calculated from the pressure corrections 
\begin{equation}
 \Delta_l(r) =-\frac{p_l(r)}{dp^{(0)}(r)/dr}~. \nonumber  
\label{ds} 
\end{equation}
$l\in \{0,2\}$ is the polynomial index in the angular expansion. 
The function $R(R_0,\theta)$ is the distance of the star surface from the
center of the configuration in the direction with the angle $\theta$ to the 
polar axis. In particular, we can define the equatorial radius 
$R_e=R(R_0,\theta=\pi/2)$ and the polar radius $R_p=R(R_0,\theta=0)$ and the 
eccentricity $\epsilon=\sqrt{1-(R_p/R_e)^2}$.

The correction to the momentum of inertia $\Delta I(r)=I(r)-I_0(r)$ can be
represented in the form
 \begin{equation}
\Delta I=\Delta I_{\rm Redist.}+\Delta I_{\rm Shape}
+\Delta I_{\rm Field}+\Delta I_{\rm Rotation}~.
\end{equation}
The first three contributions can be expressed by integrals of the angular
averaged modifications of the matter distribution, the shape of the 
configuration and the gravitational fields, in the form
\begin{equation}
\Delta I_{\alpha }=\int_0^{I_0(R_0)}dI_0(r)[W_0^{(\alpha
)}(r)-{W_2^{(\alpha )}(r)}/{5}]~, 
\label{di}
\end{equation}
where
\begin{eqnarray}
W_l^{\rm (Field)}(r)
&=&\left(\frac{\Omega}{ \Omega_0}\right)^2 
\left\{2U_l(r)-[f_l(r)+\Phi _l(r)]/{2}\right\},\\
W_l^{\rm (Shape)}(r)
&=&\left(\frac{\Omega}{\Omega_0}\right)^2 
\frac{d~\Delta_l(r)}{dr}~,\\
W_l^{\rm (Redist.)}(r)
&=&\left(\frac{\Omega}{ \Omega_0}\right)^2
\frac{p_l(r)+\varepsilon_l(r)}{p^{(0)}(r)+\varepsilon ^{(0)}(r)}~,
\end{eqnarray}
respectively, which have to be determined from the Eq. (\ref{einstein}) in 
second order approximation. 
The contribution of the change of the rotational energy to the moment of 
inertia 
\begin{equation}
\Delta I_{\rm Rotation}=\frac 45\int_0^{I_0(R_0)}dI_0(r) 
\left[r^2\bar \omega ^2(r)e^{-\nu _0(r)}\right]~.
\end{equation}
includes the frame dragging contribution.
In this expansion we neglect the influence of the change of the frame dragging
frequency, since it corresponds to the next order of the perturbative 
expansion $\sim O(\Omega^3)$.
For a more detailed description of the solutions of the field equations
in the $\sim O(\Omega^2)$ approximation we refer to the works of \cite{hartle} 
as well as \cite{chubarian}. 

\section{Model EOS with deconfinement phase transition}
\label{sec:eos}

For the investigation of the deconfinement phase transition expected to occur 
in neutron star matter at densities above the nuclear saturation density
$n_0=0.16~{\rm fm}^{-3}$ several approaches to quark confinement dynamics
have been discussed, see e.g. \cite{bkt,b+99,drago} which lead to interesting
conclusions for the properties of quark matter at high densities.
Most of the approaches to quark deconfinement in neutron star matter, however,
use a thermodynamical bag-model for the 
quark matter and employ a standard two-phase description of the 
equation of state (EOS) where the hadronic phase and the quark matter phase 
are modeled separately and the resulting EOS is obtained by imposing Gibbs'
conditions for phase equilibrium with the constraint that baryon number as 
well as electric charge of the system are conserved (Glendenning 1992, 1997). 
Since the focus of our work is the elucidation of qualitative features of 
signals for a possible deconfinement transition in the pulsar timing, we will
consider here such a rather standard, phenomenological model for an EOS with 
deconfinement transition. 

The total pressure $p(\{\mu_i\},T)$ as a thermodynamical potential
is given by
\begin{eqnarray}
\label{pres}
p(\{\mu_i\},T)&=&(1-\chi)p_H(\{\mu_h\},T) \nonumber\\
&& + \chi p_Q(\{\mu_q\},T)+p_L(\{\mu_l\},T)~~,
\end{eqnarray}
where
\begin{eqnarray}
p_H(\{\mu_h\},T)&=&\sum_{h=n,p} p^{id}_h(\mu_h^*,T;m_h^*)\nonumber\\
&&+\frac 12 (g_\omega \bar{\omega}_0)^2-\frac 12(g_\sigma \bar{\sigma})^2
\end{eqnarray}
is the EOS of the relativistic $\sigma-\omega$ mean-field model 
(Walecka model) for nuclear matter (Walecka 1974, Kapusta 1989), where
the masses and chemical potentials 
have to be renormalized by the mean-values of the $\sigma-$ and $\omega-$
fields  
$m_h^*=m_h-g_\sigma \bar \sigma$, $\mu_h^*=\mu_h-g_\omega \bar \omega_0$.
The pressure for two-flavor quark matter within a bag model EOS is given
by
\begin{equation}
p_Q(\{\mu_q\},T)=\sum_{q=u,d} p^{id}_q(\mu_q,T;m_q)-B
\end{equation}
where $B$ denotes the phenomenological bag pressure that
enforces quark confinement and the transition to nuclear matter at low 
densities. For our numerical analyses in the present work, we assume here a 
value of $B=75 ~{\rm MeV fm}^{-3}$ which allows us, e.g., to discuss a neutron 
star of $1.4$ solar masses with an extended quark matter core.

In a neutron star, these phases of strongly interacting matter are in 
$\beta-$ equilibrium with electrons and muons which contribute to the 
pressure balance with 
\begin{equation}
p_L(\{\mu_l\},T)=\sum_{l=e^-,\mu^-} p^{id}_l(\mu_l,T;m_l)~~.
\end{equation}
In the above expressions 
$p^{id}_i(\mu_i,T;m_i)=p^{-}_i(\mu_i,T;m_i)+p^{+}_i(\mu_i,T;m_i)$ 
is the partial pressure of the Fermion species $i$ as a sum of particle and
antiparticle contributions defined by 
\begin{equation}
p^{\pm}_i(\mu_i,T;m_i)=\gamma_i\int_0^\infty\frac{d^3k}{(2 \pi)^3}
T \ln[1+\exp(x^{\pm}(k))]~~,
\end{equation}
where $x^{\pm}(k)=\sqrt{k^2+m_i^2}/T\pm \mu/T$.
All the other thermodynamic quantities of interest can be derived from the 
pressure (\ref{pres}) as, e.g., the partial densities of the species
\begin{equation}
n_j(\{\mu_i\},T)=\frac{\partial p(\{\mu_i\},T)}{\partial \mu_j}~.
\end{equation}

The chemical equilibrium due to the direct and inverse $\beta$- decay 
processes imposes additional constraints on the values of the chemical 
potentials of leptonic and baryonic species (Glendenning 1997, Sahakian 1995). 
Only two independent chemical potentials remain according to the 
corresponding two conserved charges of the system, the total baryon number 
$N_B$ as well as electrical charge $Q$ 
\begin{eqnarray}
\label{bn}
n_B&=&\frac{N_B}{V}=
(1-\chi)n_H(\{\mu_h\},T) + 
\chi n_Q(\{\mu_q\},T)\\
0&=&\frac{Q}{V}=(1-\chi)q_H(\{\mu_h\},T) + 
\chi q_Q(\{\mu_q\},T)\nonumber\\
&& ~~~~~+q_L(\{\mu_l\},T)~.
\label{charge}
\end{eqnarray}
The deconfinement transition is obtained
following the construction  
which obeys the global conservation laws 
and allows one to find the volume fraction of the quark matter phase 
$\chi=V_Q/V$ in the mixed phase where $p_H(\{\mu_h\},T) = p_Q(\{\mu_q\},T)$, 
so that at given $n_B$ and $T$ the total pressure under the conditions 
(\ref{bn}), (\ref{charge}) is a maximum, see  Glendenning (1992, 1997).

\begin{figure}[bht]
\psfig{figure=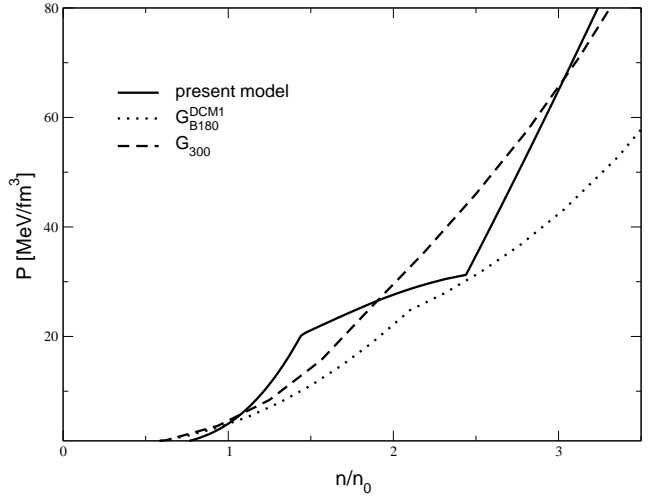,width=8.8cm,angle=-90}
\caption
{Model equation of state for the pressure of hybrid star matter in 
$\beta$-equilibrium as a function of the baryon number density. 
The hadronic equation of state is a relativistic mean-field model 
($\sigma-\omega$), the quark matter one is a 
two-flavor bag model with $B=75$ MeV fm$^{-3}$.
For comparison the relativistic mean-field EoS of \protect\cite{glen89} 
including $\rho$ mesons, hyperons and muons 
(incompressibility $K=300$ MeV, dotted line) and that of \protect\cite{glen} 
including a deconfinement transition to three-flavor quark matter in a bag 
model with $B=180$ MeV$^4$ (dashed line) are shown.
\label{fig1}}
\end{figure}

In Fig. \ref{fig1} we show the model EoS with deconfinement transition as
described above. Note that in the density region of the phase transition 
there is a monotonous increase of the pressure which gives rise to an extended 
mixed phase region in the compact star after solution of the equations of 
hydrodynamic stability (\ref{TOV}).
For comparison, the relativistic mean-field EoS of \protect\cite{glen89} 
including $\varrho$ mesons, hyperons and muons 
(incompressibility $K=300$ MeV, dotted line) and that of \protect\cite{glen} 
including a deconfinement transition to three-flavor quark matter in a bag 
model with $B=180$ MeV$^4$ are shown, see also the monographs by 
\cite{gbook} and \cite{fwbook}.

\begin{figure}[bht]
\psfig{figure=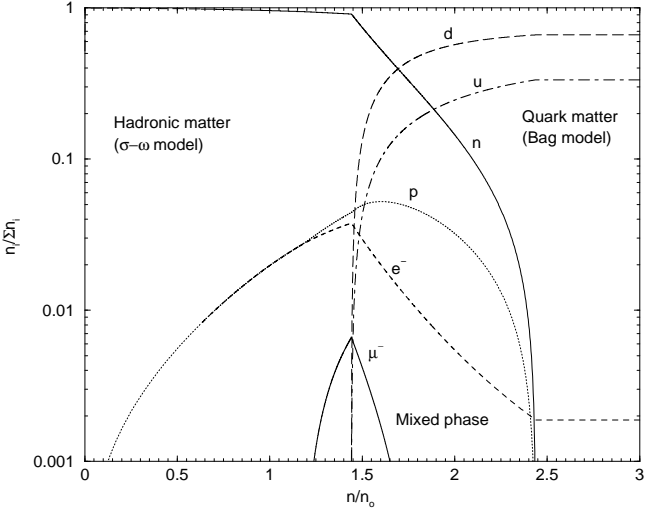,width=8.6cm,angle=-90}
\caption
{Composition of hybrid star matter in $\beta$-equilibrium as a 
function of baryon number density for the model EoS used in this work. 
\label{fig2}}
\end{figure}

In Fig. \ref{fig2} we show the composition of the hybrid star matter as a 
function of the total baryon density at $T=0$. 
Solving the Tolman-Oppenheimer-Volkoff equations (\ref{TOV})-
%, (\ref{mass}) 
(\ref{phs}) for the hydrodynamical 
equilibrium of static spherically symmetric relativistic stars with the 
above defined EOS, we find that a configuration at the stability limit 
could have a quark matter core with a radius as large as $\sim 75 \% $ 
of the stars radius.

What implications this phase transition for rotating star configurations 
might have will be investigated in the next section by applying the method 
developed in Sect. \ref{sec:GR} for the above EOS.

\section{Results and discussion}
\label{sec:result}

\begin{figure}[bht]
\psfig{figure=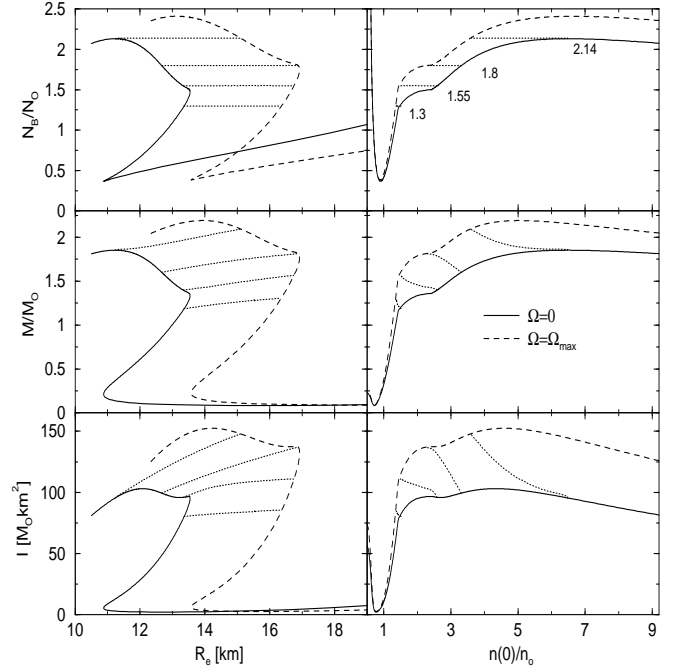,width=8.6cm,height=9cm,angle=-90}
\caption
{Baryon number $N$, mass $M$, and moment of inertia $I$ as a function
of the equatorial radius (left panels) and the central density 
(right panels) for neutron star configurations with a deconfinement 
phase transition. 
The solid curves correspond to static configurations, the dashed ones
to those with maximum angular velocity $\Omega_{\rm K}$. 
The lines between both extremal cases connect configurations with the
same total baryon number $N_B/N_{\odot}=1.3, 1.55, 1.8, 2.14$.
\label{fig3}}
\end{figure}

The results for the stability of rotating neutron star configurations with
possible deconfinement phase transition according to the EOS described in the 
previous section are shown in Fig. \ref{fig3}, where the total baryon number, 
the total mass, and the moment of inertia are given as functions of the 
equatorial radius (left panels) and in dependence on the central baryon number
density (right panels) for static stars (solid lines) as well as for stars 
rotating with the maximum angular velocity $\Omega_{\rm max}$ (dashed lines). 

The dotted lines connect configurations with fixed total baryon numbers
$N_B/N_\odot = 1.3, 1.55, 1.8, 2.14$ and it becomes apparent that the rotating
configurations are less compact than the static ones. They have larger masses,
radii and momenta of inertia at less central density such that for suitably 
chosen configurations a deconfinement transition in the interior can occur 
upon spin-down.
 
\begin{figure}[bht]
\psfig{figure=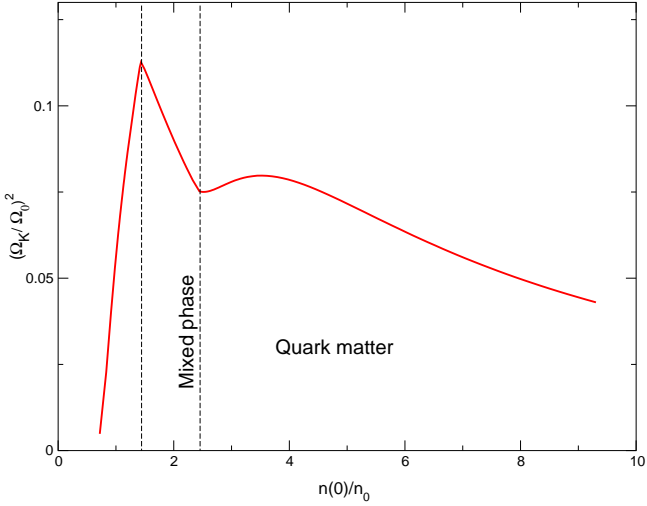,width=8.8cm,angle=-90}
\caption
{Expansion parameter for the maximally 
attainable rotation frequencies of stationary rotating objects 
$(\Omega_K/\Omega_0)^2$ 
as a function of the central density characterizing the configuration.
Vertical dashed lines indicate the density band for which a mixed phase 
occurs, see Fig. \ref{fig2}.
\label{fig4}}
\end{figure}

In order to demonstrate the consistency of our perturbative approach, we show 
in Fig. \ref{fig4} the values of the expansion parameter for the maximally 
attainable rotation frequencies of stationary rotating objects 
($\Omega_K/\Omega_0$) 
as a function of the central density characterizing the configuration. 

\begin{figure}[bht]
\psfig{figure=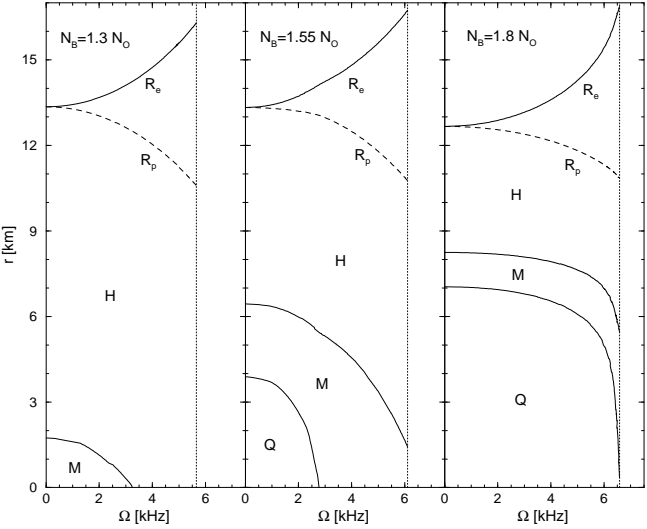,width=8.6cm,angle=-90}
\caption
{Phase structure of rotating hybrid stars in equatorial 
direction in dependence of the angular velocity $\Omega$ for stars 
with different total baryon number: $N_B/N_{\odot}=1.3, 1.55, 1.8$.
Dahsed vertical lines indicate the maximum frequency $\Omega_K$ for 
stationary rotation.
\label{fig5}}
\end{figure}

In Fig. \ref{fig5} we show the critical regions of the phase transition in
the inner structure of the star configuration as well as the
equatorial and polar radii in the plane of angular velocity $\Omega$
versus distance from the center of the star. 
It is obvious that with the increase of the
angular velocity the star is deforming its shape. 
The maximal eccentricities of the configurations with $N_B=1.3~N_\odot$,  
$N_B=1.55~N_\odot$ and $N_B=1.8~N_\odot$ are 
$\epsilon(\Omega_K)=0.7603$, 
$\epsilon(\Omega_K)=0.7655$ and 
$\epsilon(\Omega_K)=0.7659$, respectively. 
Due to the changes of the central density the quark core could disappear above
a critical angular velocity. 
 
\begin{figure}[bht]
\psfig{figure=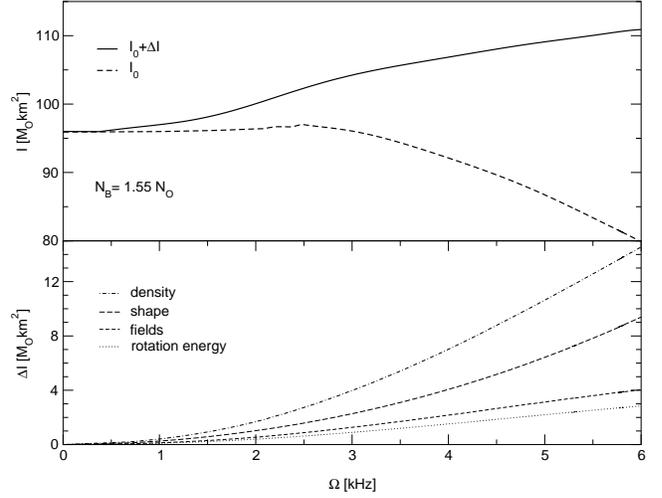,width=8.8cm,angle=-90}
\caption
{Contributions to the dependence of the moment of inertia 
on the angular velocity.
\label{fig6}}
\end{figure}

In Fig. \ref{fig6} we display the dependence of the moment of
inertia on the angular velocity for configurations with the same total 
baryon number $N_B=1.55~N_\odot$ together with the different
contributions to the total change of the moment of inertia. 
As it is shown the most important contributions come from the mass 
redistribution and the shape deformation. 
The relativistic contributions due to field and rotational energy are less 
important. 
In the same Fig. \ref{fig6} we show the decrease of the spherical
moment of inertia due to the decrease of the central density for high angular
velocities which tends to partially compensate the further increase of the 
total moment of inertia for large $\Omega$. There is no dramatic change in the
slope of $I(\Omega)$ at $\Omega_{\rm crit}=2.77$ kHz.

\begin{figure}[bht]
\psfig{figure=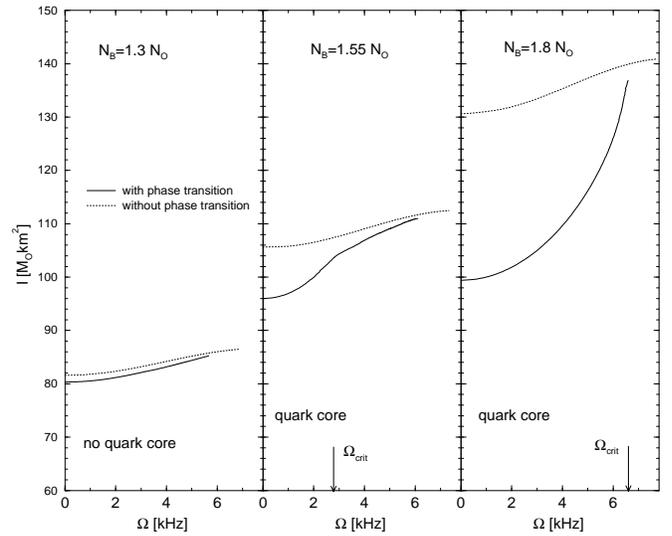,width=8.6cm,angle=-90}
\caption
{Moment of inertia as a function of angular velocity with
(solid lines) and without (dashed lines) deconfinement phase 
transition for fixed total baryon number $N_B/N_{\odot}=1.3$ 
(left panel), $N_B/N_{\odot}=1.55$  (middle panel), 
$N_B/N_{\odot}=1.8$ (right panel).
\label{fig7}}
\end{figure}

Fig. \ref{fig7} shows the dependence of the
moment of inertia as a function of the angular velocity. It is demonstrated
that the behavior of $I(\Omega)$ for a given total number of
baryons $N_B$ strongly depends on the presence of a pure quark matter core in 
the center of the star. 
If the core does already exist or it does not appear when the 
angular velocity increases up to the maximum value $\Omega_{\rm max}$ 
then the second order derivative of the moment of inertia $I(\Omega)$ does 
not change its sign. 
For the configuration with $N_B=1.55~N_\odot$ the critical value for the 
occurrence of the sign change is $\Omega_{\rm crit}= 2.77$ kHz while for 
$N_B=1.8~N_\odot$ it is close to $\Omega_K= 6.6$ kHz. 

In order to point out possible observable consequences of such a 
characteristic behaviour of $I(\Omega, N_B)$ we consider two possible 
scenarios for changes in the pulsar timing:
(A) dipole radiation and the resulting dependence of the braking index on the 
angular velocity as suggested by \cite{frido} and
(B) mass accretion onto rapidly rotating neutron stars.

\subsection{Magnetic dipole radiation}
Due to the energy loss by magnetic dipole radiation plus emission of 
electron-positron wind the star has to spin-down and the 
resulting change of the angular velocity can be parametrized by a power law 
\begin{equation}
\frac{\dot \Omega }{\Omega}= - K \Omega^{n(\Omega)-1},  
\end{equation}
where $K$ is a constant and $n(\Omega)$ is the braking index 
\begin{equation}
\label{bi}
n(\Omega)=\frac{\ddot{\Omega}~\Omega}{\dot\Omega^2}= 3 - 
\frac{3 I'\Omega+I''\Omega^2 }{2 I+I'\Omega},  
\end{equation}
where we used the notation 
$I'=(\partial I(\Omega, N_B)/ \partial \Omega)|_{N_B={\rm const}}$, 
with the corresponding definition of $I''$ (see also 
Glendenning et al. 1997; Grigorian et al. 1999).

\begin{figure}[bht]
\psfig{figure=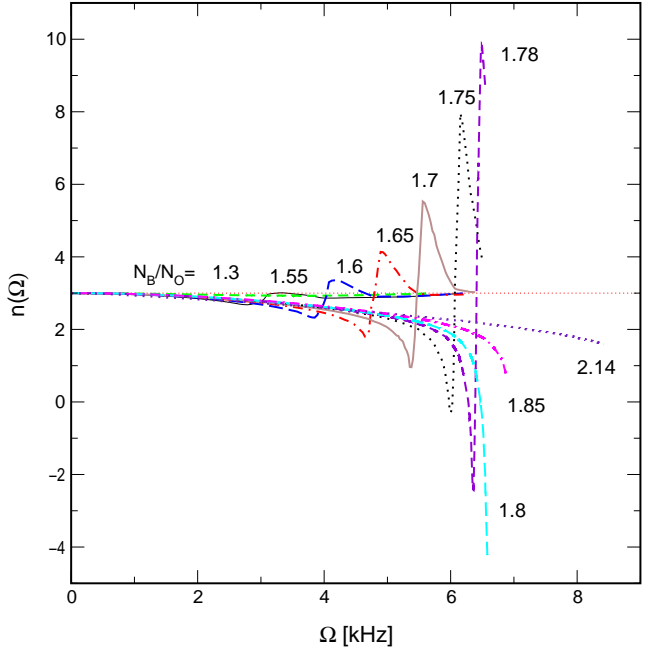,width=8.8cm,angle=0}
\caption
{Braking index due to dipole radiation from fastly rotating 
isolated pulsars as a function of the angular velocity.
The minima of $n(\Omega)$ indicate the appearance/ disappearance 
of quark matter cores.
\label{fig8}}
\end{figure}

In Fig. \ref{fig8} we display the result for the braking index $n(\Omega)$ 
for a set of configurations with fixed total baryon numbers ranging from 
$N_B=1.55~N_\odot$ up to $N_B=1.9~N_\odot$, the region where during the 
spin-down evolution a quark matter core could occur for our model EOS.
We observe that only for configurations within the interval of total 
baryon numbers $1.4\le N_B/N_\odot\le 1.9$ a quark 
matter core occurs during the spin-down as a consequence of the increasing 
central density (see also Fig. \ref{fig5}) and the braking index shows 
variations.  
The critical angular velocity $\Omega_{\rm crit}(N_B)$ for the appearance of 
a  quark matter core can be found from the minimum of the braking index Eq. 
(\ref{bi}).
As can be seen from Fig. \ref{fig8}, all configurations with a quark 
matter core have braking indices $n(\Omega) < 3$ and braking indices 
significantly larger than 3 can be considered as precursors of the 
deconfinement transition. 
The magnitude of the jump in  $n(\Omega)$ during the transition to 
the quark core regime is a measure for the size of the quark core. 
It would even be sufficient to observe the maximum of the braking index 
$n_{\rm max}$ in order to infer not only the onset of deconfinement 
($\Omega_{\rm max}$) but also the size of the quark core to be 
developed during further spin-down from the maximum deviation 
$\delta n = n_{\rm max} - 3$ of the braking index.
For the model EOS we used a significant enhancement of the braking 
index which does only occur for pulsars with periods $P<1.5$ ms (corresponding 
to $\Omega>4$ kHz), which have not yet been observed in nature.
Thus the signal seems to be a weak one for most of the possible candidate 
pulsars.
However, this statement is model dependent since, e.g., for the model EOS 
used by \cite{frido}, which includes the 
strangeness degree of freedom, a more dramatic signal at lower spin 
frequencies has been reported.
Therefore, a more complete investigation of the braking index for a set of 
realistic EOS should be performed.

\subsection{Mass accretion}
A higher spin-down rate than in isolated pulsars might be possible for 
rotating neutron stars with mass accretion. 
In that case, at high rotation frequency the angular momentum transfer from 
accreting matter and the influence of 
magnetic fields can then be neglected (Shapiro \& Teukolsky 1983)
so that the evolution of the angular velocity is determined by the 
dependence of the moment of inertia on the total mass, i.e. baryon number,
\begin{equation}
\frac{\dot \Omega}{\Omega}= -  
\left(\frac{N_B}{I} \frac{{\rm d}I}{{\rm d}N_B}\right)\bigg|_{J={\rm const}}
\frac{\dot N_B}{N_B}~,
\end{equation}
where $J=I~\Omega={\rm const}$ has been assumed.

\begin{figure}[bht]
\psfig{figure=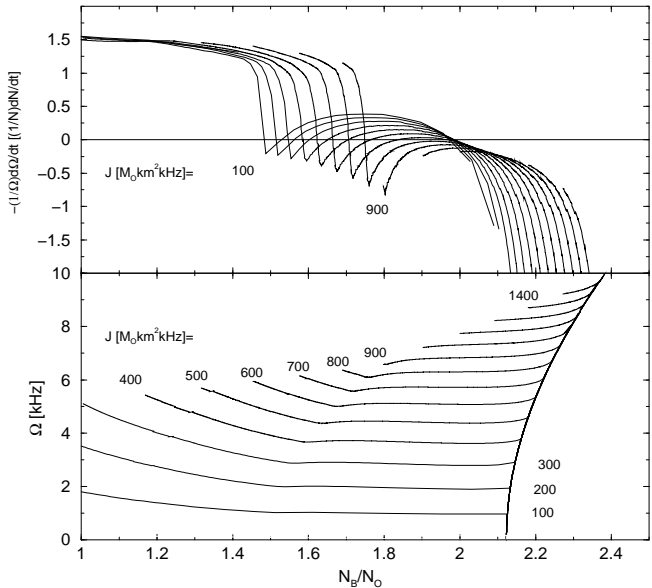,width=8.6cm,angle=-90}
\caption
{Total baryon number dependence of the spin-down rate 
$\dot \Omega/\Omega$ in units of the baryon number accretion rate
($\dot N_B/N_B$) and the corresponding angular velocity (lower panel)
for different (conserved) total angular momenta 
$J[{\rm M}_\odot {\rm km}^2 {\rm kHz}]=100, 200, \dots, 1400 $.
The suggested signal for a deconfinement transition in rapidly 
rotating neutron stars with baryon number accretion is a transition 
from a spin-down to a spin-up regime, i.e. a zero in the spin-down 
rate.
\label{fig9}}
\end{figure}

In Fig. \ref{fig9} we consider the change of the pulsar timing 
due to mass accretion with a constant accretion rate $\dot N_B/N_B$ 
for fixed total angular momentum as a function of the total baryon number. 
Here the change from spin-down to spin-up behaviour during the pulsar 
evolution signals the deconfinement transition. 
When the pulsar has developed a quark matter core then the change of the 
moment of inertia due to further mass accretion is negligible and has no 
longer influence on the pulsar timing. 
However, in real systems the transfer of angular momentum from the accreting 
matter can lead to a spin-up already. 
Then the transition to the quark matter core regime should be observable as an 
increase in the spin-up rate.

It will be interesting to investigate in the future whether e.g. 
low-mass X-ray binaries (LMXBs) with mass accretion might be discussed as 
possible candidates for rapidly rotating neutron stars for which consequences 
of the transition to the quark core regime due to mass accretion might be 
detected. 
Recently, quasi-periodic brightness oscillations (QPO's) with frequences up to 
$\sim 1200$ Hz have been observed (Lamb et al. 1998) which entail new mass and 
radius constraints for compact objects.
Note in this context that the assumption of a deconfined matter interior 
of the compact star in some LMXBs as e.g. SAX J1808.4-3658 (Li et al. 1999)
seems to be more consistent than that of an ordinary hadronic matter interior. 

\section{Conclusions}

It has been demonstrated, through the example of the deconfinement transition 
from hadronic to quark matter, that the rotational characteristics of neutron 
stars (braking index, spin-down rate) are sensitive to changes of their inner 
structure and can thus be investigated in order to detect structural phase
transitions.
  
The theoretical basis for the present work was a perturbation method for the 
solution of the Einstein equations for axial symmetry which allows us to 
calculate the contribution of different
rotational effects to the change of the moment of inertia.
This quantity can be used
as a tool for the investigation of the changes in the rotation timing for
different scenarios of the neutron star evolution.

The deviation of the braking index from the value $n=3$ (magnetic dipole 
radiation)  as a function of the angular velocity has been suggested as a 
possible signal for the deconfinement transition and the occurrence of a quark 
matter core in pulsars. 
We have reinvestigated this signal within our approach and could show that 
the magnitude of this deviation is correlated with the size of the quark core,
since the influence of the $Ae$ phase crust on such processes is negligible. 

For neutron stars with mass accretion, we have suggested that under the 
assumption of total angular momentum conservation a flip from spin-down to 
spin-up behaviour signals the appearance of a quark matter core.
A more detailed investigation is necessary to identify possible 
candidates of rotating compact objects with mass accretion (see e.g. 
\cite{rxte}) for which the suggested deconfinement signal could be relevant. 
  
Although the quantitative details of the reported deconfinement signals are 
quite model-dependent and might change when one uses more realistic equations 
of state, e.g. including the strangeness degree of freedom 
(Glendenning et al. 1997, Weber 1999), the relation between the magnitude of 
the effect and the 
size of the quark core which has been found here is expected to be 
model-independent and should be confirmed by subsequent studies.     

\section*{Acknowledgments}
We thank A. Drago, F. Weber and N. Glendenning for their comments on this 
work.
D.B. acknowledges discussions during the workshop on ``Understanding 
Deconfinement in QCD'' at the ECT* in Trento.
Our thanks go to R. Avakian, A. Sedrakian and D. Sedrakian for their 
stimulating interest in this work and for their useful remarks.
The work of E.C. and H.G. was supported by the {Volkswagen Stiftung} 
under grant No. I/71 226. G.P. received support from the {Deutscher 
Akademischer Austauschdienst (DAAD)} and H.G. acknowledges a stipend from
the {Deutsche Forschungsgemeinschaft} (DFG), grant no. 436 ARM 17/1/00.

\end{document}